\documentclass[fleqn,usenatbib]{mnras}

\usepackage[T1]{fontenc}
\usepackage{ae,aecompl}
\usepackage{graphicx}
\usepackage{multicol}
\usepackage{float}

\usepackage{graphicx}	
\usepackage{amsmath}	
\usepackage{amssymb}	
\usepackage{threeparttable}
\usepackage{subfigure}
\usepackage{multirow}
\usepackage{multicol}
\usepackage{booktabs}
\usepackage{makecell}

\newcommand{\Lya}{\ensuremath{\hbox{Ly}\alpha~}}

\newcommand{\Ka}{\ensuremath{\hbox{K}\alpha~}}
\newcommand{\Kb}{\ensuremath{\hbox{K}\beta~}}
\newcommand{\FeXXV}{\mbox{{Fe\,{\sevensize XXV}~}}}
\newcommand{\FeXXVI}{\mbox{{Fe\,{\sevensize XXVI}~}}}
\def\cha{{\it Chandra~}}
\def\fer{{\it Fermi~}}

\title[Accretion disk in Vela X-1]{Spectral evidence of an accretion disk in wind-fed
X-ray pulsar Vela X-1 during an unusual spin-up period}

\author[Z. X. Liao et al.]{
Zhenxuan Liao$^{1,2}$,
Jiren Liu$^{1}$\thanks{E-mail: jirenliu@nao.cas.cn},
Xueying Zheng$^{1,2}$,
and Lijun Gou$^{1,2}$
\\
$^{1}$Key Laboratory for Computational Astrophysics, National Astronomical Observatory, Chinese Academy of Sciences, \\~~Datun Road 20A, Beijing 100012, People's Republic of China\\
$^{2}$School of Astronomy and Space Science, University of Chinese Academy of Sciences, Beijing 100049, People's Republic of China\\
}

\date{Accepted XXX. Received YYY; in original form ZZZ}

\pubyear{2020}

\begin{document}
\label{firstpage}
\pagerange{\pageref{firstpage}--\pageref{lastpage}}
\maketitle

\begin{abstract}
In classical supergiant X-ray binaries (SgXBs), the Bondi-Hoyle-Lyttleton wind accretion was usually assumed,
and the angular momentum transport to the accretors is inefficient. The observed spin-up/spin-down behavior of 
the neutron star in SgXBs is not well understood. In this paper, we report an extended low state of Vela X-1 (at 
orbital phases 0.16--0.2), lasting for at least 30\,ks, observed with \cha during the onset of an unusual
spin-up period. During this low state, the continuum fluxes dropped by a factor of 10 compared to the preceding 
flare period, and the continuum pulsation almost disappeared. Meanwhile, the Fe \Ka fluxes of the low state
were similar to the preceding flare period, leading to an Fe \Ka equivalent width (EW) of 0.6\,keV, as high as the
Fe \Ka EW during the eclipse phase of Vela X-1. Both the pulsation cessation and the high Fe \Ka EW indicate an 
axisymmetric structure with a column density larger than $10^{24}$\,cm$^{-2}$ on a spatial scale of the 
accretion radius of Vela X-1. These phenomena are consistent with the existence of an accretion disk that 
leads to the following spin-up of Vela X-1. It indicates that disk accretion, although not always, does 
occur in classical wind-fed SgXBs.

\end{abstract}




\section{Introduction}
\label{sec:intro}
In supergiant X-ray binaries (SgXBs), the X-ray radiation is powered by accretion of massive stellar wind to the compact neutron star or black hole. The accretion processes involve many physical factors, such as the clumpy wind, the strong magnetic field of the neutron star, the dynamical interaction, and the photoionization, all of which make the accretion processes of wind-fed SgXBs not fully understood \citep[for a recent review, see][]{Mar17}. 
In classical SgXBs, like Vela X-1, the Bondi-Hoyle-Lyttleton wind accretion was usually assumed \citep{DO73}, and the angular momentum transport to the neutron star is inefficient. Their pulse periods, however, show alternative spin-up and spin-down, superimposed on a longer trend of either spin-up or spin-down \citep[e.g.][]{Bild1997}. 
The pulse frequency variations of Vela X-1 were explained as a random walk in pulse frequency \citep{Dee1989}.
The angular momentum transport in wind-fed SgXBs was first discussed by \citet{SL76} and \citet{Wang81}, and
recently by \citet{Kar19}, among others. They found that when the binary orbit is tight and the stellar wind is slow,
the spatial gradient of density and velocity could lead to the formation of a disk.

\setcounter{figure}{1}
\begin{figure*}
    \includegraphics[width=2.3\columnwidth]{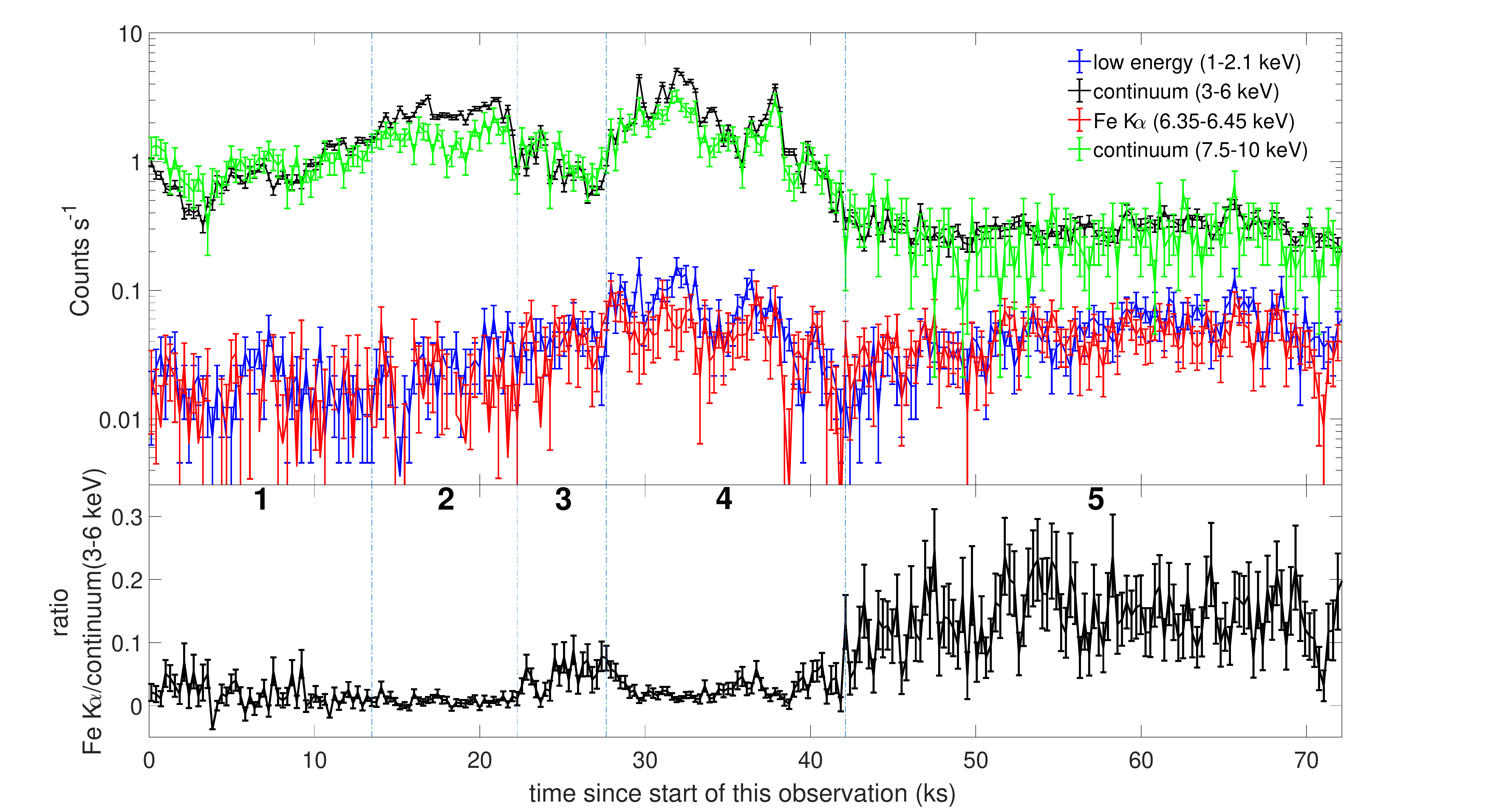}
    \caption{Light curves of the low-energy band (1--2.1\,keV), the Fe \Ka line (6.35--6.45\,keV), and the continuum within 3--6\,keV and 7.5--10\,keV band of Vela X-1 from \cha observation (top panel), along with the ratio of Fe \Ka to continuum within 3--6\,keV (bottom panel). The time resolution is 283.6 s. 
    A baseline continuum (40\% of the count rate in the 6--6.2\,keV band) has been subtracted from the Fe \Ka fluxes. For clarity, the rates of 7.5--10\,keV continuum are multiplied by a factor of 10. The individual segments analyzed later are marked as numbers of 1--5. Error bars indicate $1\sigma$ uncertainty.}
    \label{fig:open}
\end{figure*}
\setcounter{figure}{0}
\begin{figure}
    \includegraphics[angle=0, width=1.1\columnwidth]{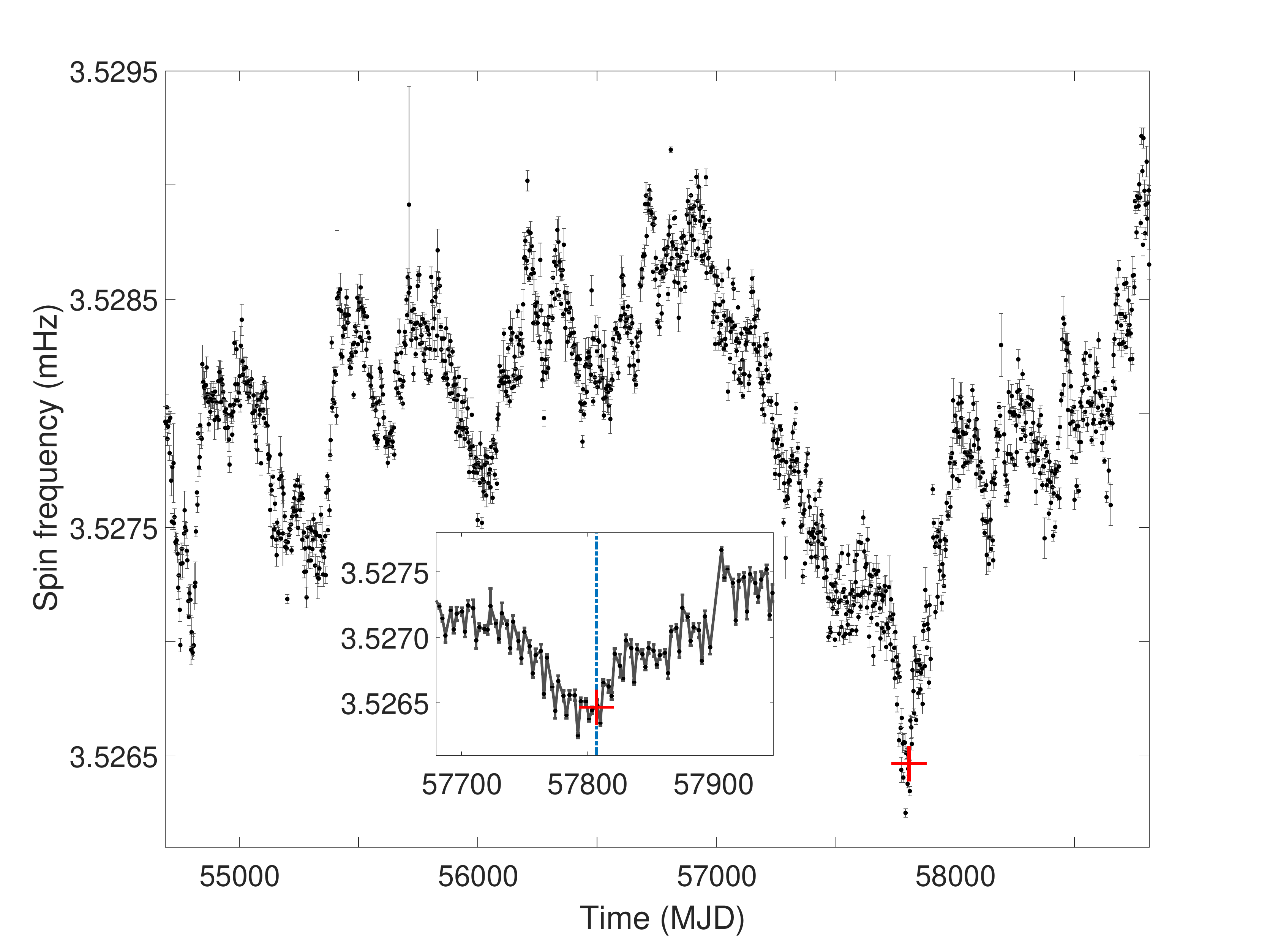}
    \caption{The pulse frequency history of Vela X-1 monitored with {\it Fermi}/GBM. Over-plotted red cross marks the
	 \cha observation taken at the onset of the spin-up event.}
    \label{fig:vx1_gbm}
\end{figure}

The formation of a disk-like structure in Vela X-1 was recently evaluated with hydrodynamic simulations by \citet{Mel19}. They found that when the wind velocity at the position of the neutron star is comparable to the orbital velocity, the wind is seriously beamed, and a disk-like structure is formed in the shocked region for a configuration of a heavy neutron star. Because Vela X-1 is an eclipsing system and the wind-captured disk is expected to form within the orbital plane, \citet{Mel19} pointed out that the disk would intercept the line-of-sight and might significantly contribute to the observed absorption column density.

Discovered in the early years of X-ray astronomy \citep{Chod1967}, Vela X-1 is one of the best studied SgXBs \citep[for a recent review, see][]{Kret2019}.
It has a pulse period of $283$\,s \citep{McC1976} and a magnetic field around
$2.6\times10^{12}\,\rm G$ \citep[e.g.][]{Kret1996}.
The optical companion, HD\,77581, is a B0.5 Ib supergiant, with a radius of $30.0\,\rm R_{\sun}$ and a mass of
$23.5\,\rm M_{\sun}$ \citep{vanKer1995}.
The binary system has an orbital period of 8.9644\,days \citep{Fala2015}, a tight orbital seperation of $53.4\,\rm
R_{\sun}$ \citep{vanKer1995}, a low eccentricity $e \approx 0.09$ \citep[e.g.][]{Boy1986}, and a high inclination
\citep[$i > 73^{\circ}$,][]{Joss1984}. The distance to Vela X-1 was estimated to be $1.9\pm0.2\,\rm kpc$ \citep{Sada1985}.

In this paper we report evidences of an accretion disk in Vela X-1 observed by \cha during the onset of an unusual
spin-up period. This \cha observation of Vela X-1 showed an extended low state, lasting for at least 30\,ks,
until the end of the observation. 
During this low state, the continuum fluxes reduced by an order of magnitude, and their pulsations almost disappeared,
while the Fe \Ka fluxes were similar to those preceding the low state, leading to an Fe \Ka 
equivalent width (EW) of $0.6$\,keV. Such a high Fe \Ka EW is only observed to appear during the eclipse phase of Vela X-1 previously.
If not specified, the quoted errors are for 90\% confidence level.

\section{Observations}
\label{sec:obs}
Thanks to the Burst and Transient Source Experiment (BATSE) on the {\it Compton} gamma ray observatory and Gamma-ray
Burst Monitor \citep[GBM,][]{Jenke2017} on the \fer gamma ray telescope, the spin history of X-ray pulsars can be continuously monitored.
The spin history of Vela X-1 monitored by {\it Fermi}/GBM is presented in Fig.~\ref{fig:vx1_gbm}. Rapid changes of
spin-up and spin-down on tens of days are clearly seen. A spin-up event around MJD 57800 (Feb. 16th, 2017)
seems unusual, in the sense that it occurred around the lowest spin
frequency monitored by GBM, with a sharp transition from spin-down to spin-up, and 
it seems to have a longest continuous spin-up period ($\sim 200$ days), although showing some flat spin periods in between. Other spin-up periods last only for about 50 days. 
This spin frequency ($\sim 3.5265$\,mHz) is also the lowest one compared with those monitored by BATSE and other early pointing observations\footnote{https://gammaray.nsstc.nasa.gov/batse/pulsar}.

Vela X-1 was happened to be observed by \cha High Energy Transmission Grating \citep[HETG,][]{Cani05} at the 
onset of this unusual spin-up period (ObsID 19953, PI Canizares, on Feb. 22nd, 2017, MJD 57806.98). The observation was taken at the orbital phase $\phi=0.107-0.203$, 
while the eclipse phase of Vela X-1 is $\phi_{\rm ecl}=0.9204-0.0899$ \citep[according to the ephemeris of][their Table 3]{Fala2015}.
The phase zero is defined as the time of mid-eclipse, $T_{\rm ecl}$, which is 0.2 days later than the 
time of mean longitude, $T_{\rm \pi/2}$.
The exposure time was 70\,ks. We downloaded the data from \cha archive and reprocessed it with {\sc tgcat} script \citep{TGCat} following the standard procedure. A narrow spatial mask is used to improve the fluxes above 7\,keV. The barycentric correction is applied with {\sc axbary} tool within CIAO software and the binary effect is corrected with the program {\it binaryCor} in Remeis ISISscripts\footnote{http://www.sternwarte.uni-erlangen.de/isis}. The time periodicity was analysed with {\it sitar\_epfold\_rate} and {\it sitar\_pfold\_rate} tools within the {\sc sitar} package\footnote{http://space.mit.edu/cxc/analysis/SITAR/distrib.html}. We only used the first order HEG data, because of its better spectral resolution and larger effective area in the Fe \Ka band than higher order data.

\section{Timing results}

\setcounter{figure}{2}
\begin{figure*}
    \subfigure{
    \label{fig:prd_search_3-6_5seg}
    \includegraphics[angle=0, width=\columnwidth]{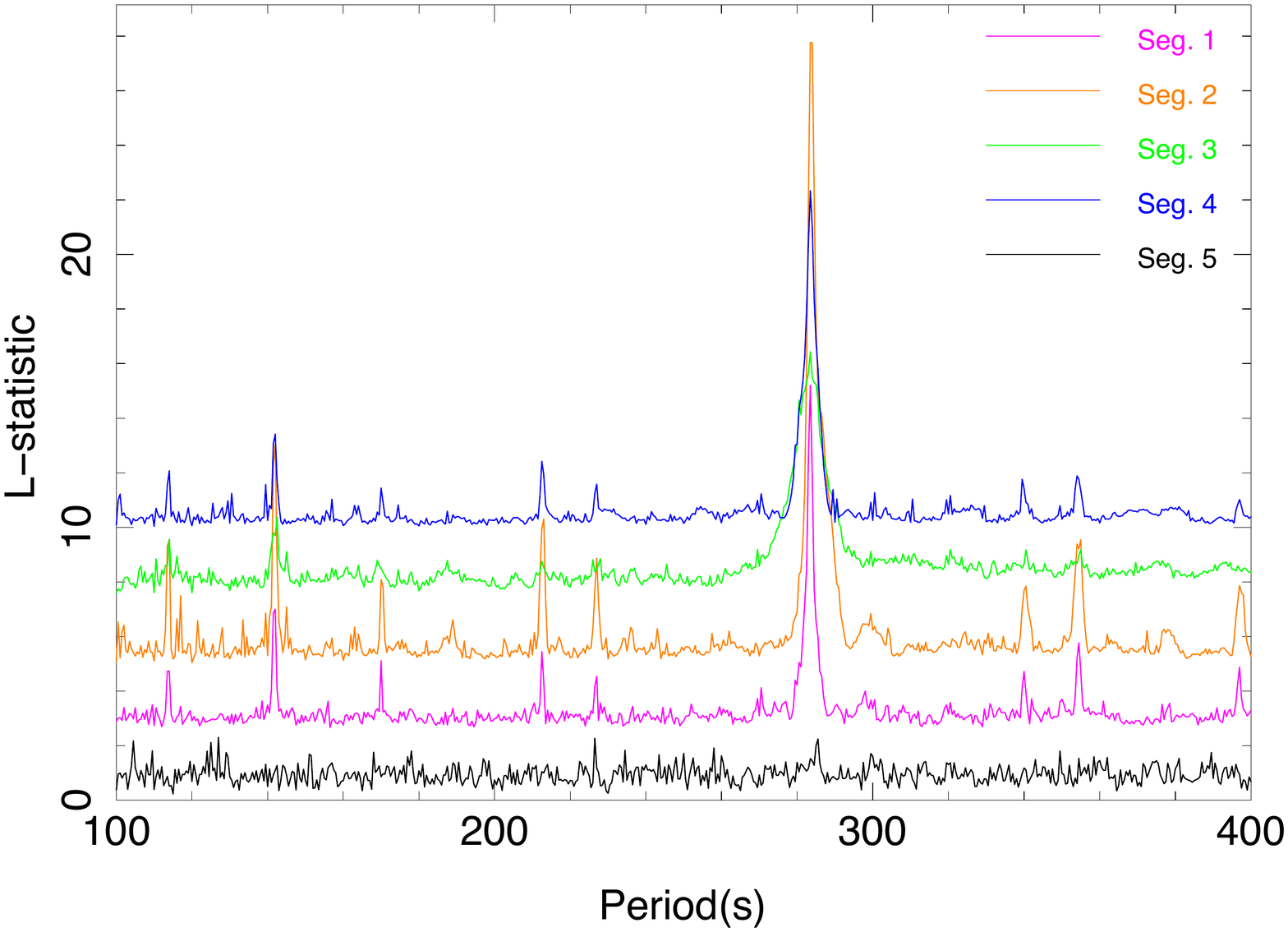}
    }
    \subfigure{
    \label{fig:prd_search_7+_5seg}
    \includegraphics[angle=0, width=\columnwidth]{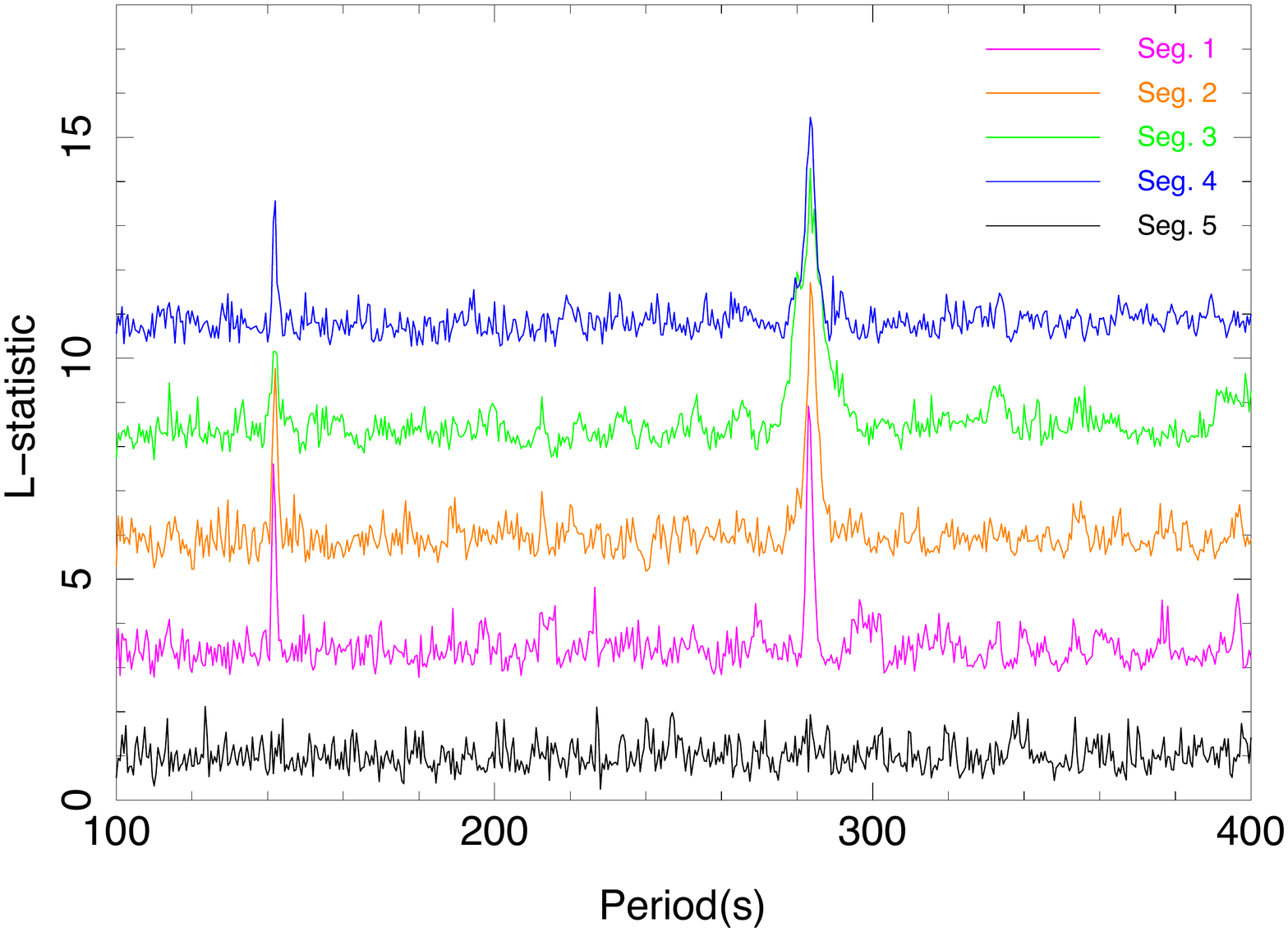}
    }
    \subfigure{
    \label{fig:prd_search_lowe_5seg}
    \includegraphics[angle=0, width=\columnwidth]{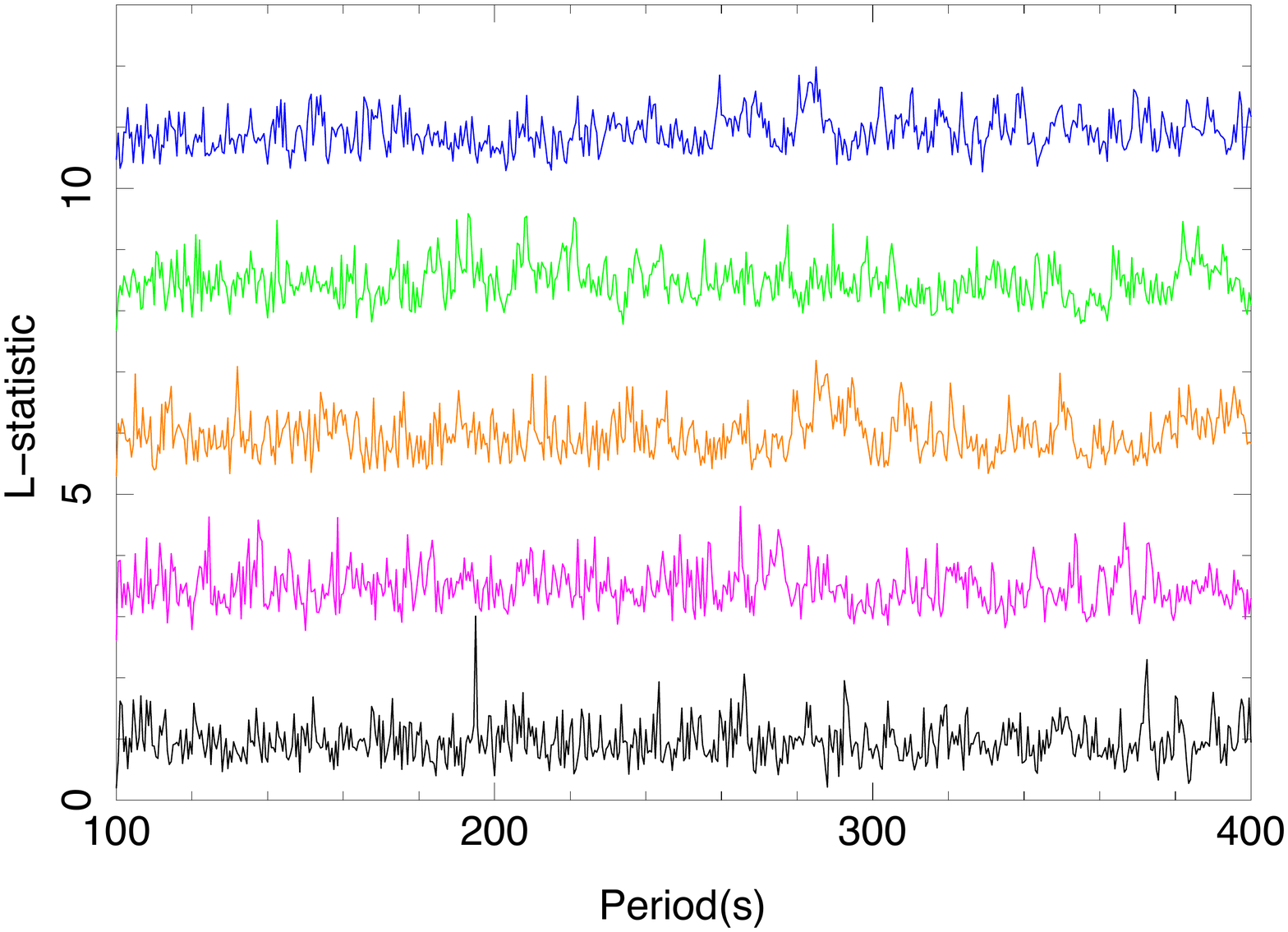}
    }
    \subfigure{
    \label{fig:prd_search_feka_5seg}
    \includegraphics[angle=0, width=\columnwidth]{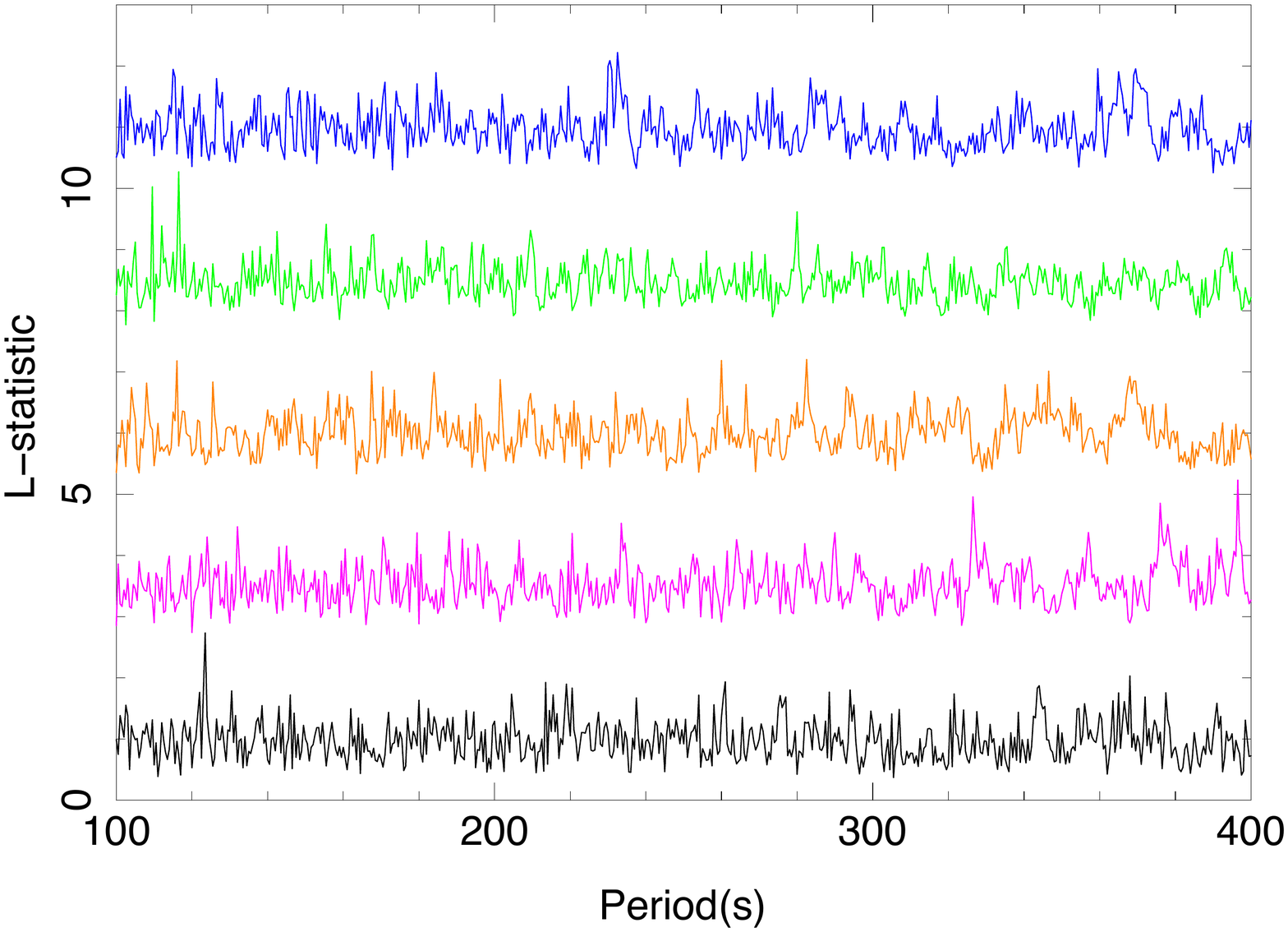}
    }
    \caption{Statistical significance of the pulsation period for the continuum (3--6\,keV, upper left panel; 7.5--10\,keV, upper right panel), the low-energy band (1--2.1\,keV, lower left panel), and the Fe \Ka line (lower right panel) during different segments. The curves of Seg.\,1-4 are shifted upwards by 2.5, 5.0, 7.5, 10.0, respectively. 
	 The color scheme used in the lower panel is the same as the upper panel.}
    \label{fig:prd_search_5seg}
\end{figure*}
\begin{figure*}
    \includegraphics[angle=0, width=\columnwidth]{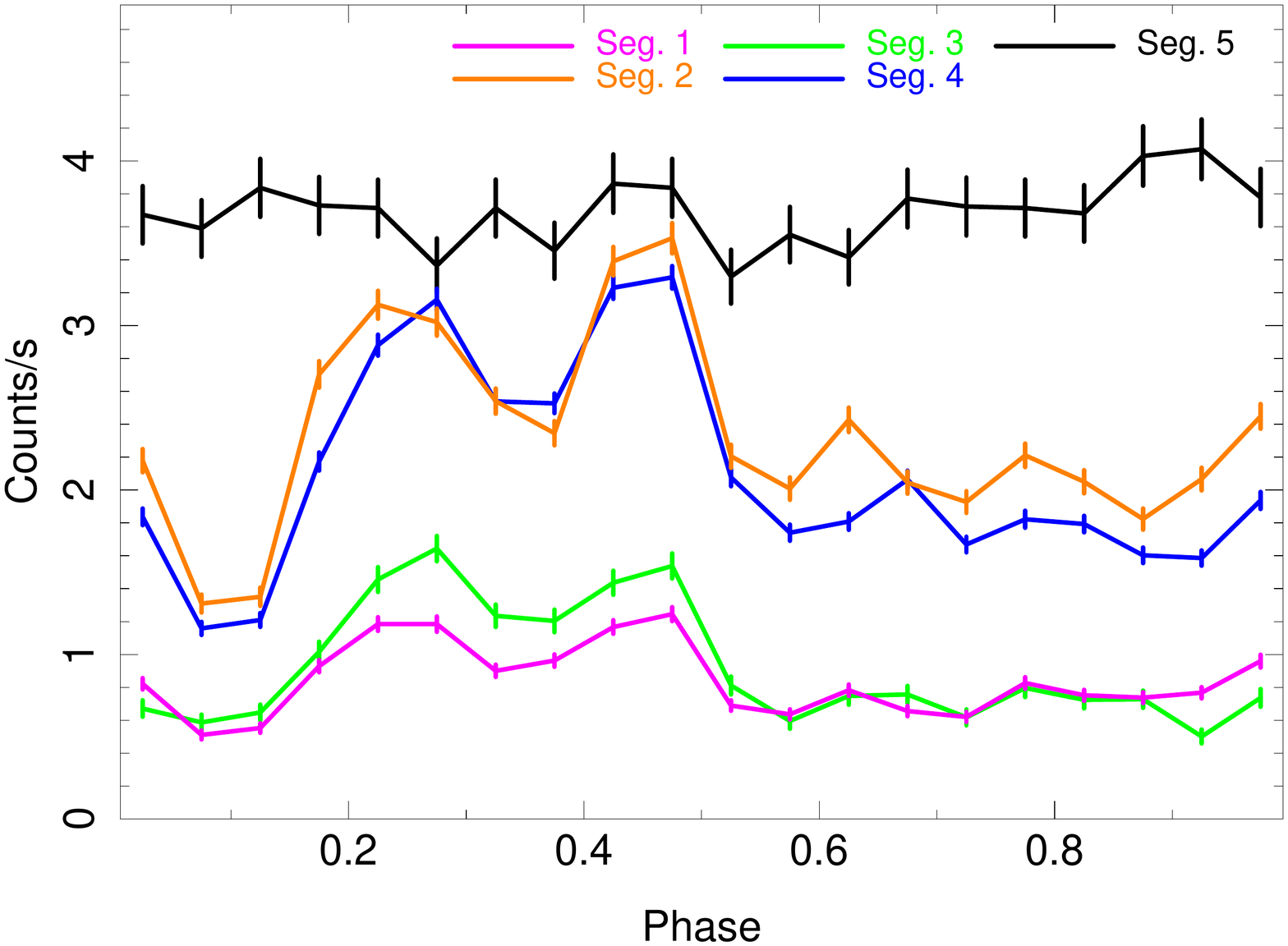}
    \includegraphics[angle=0, width=\columnwidth]{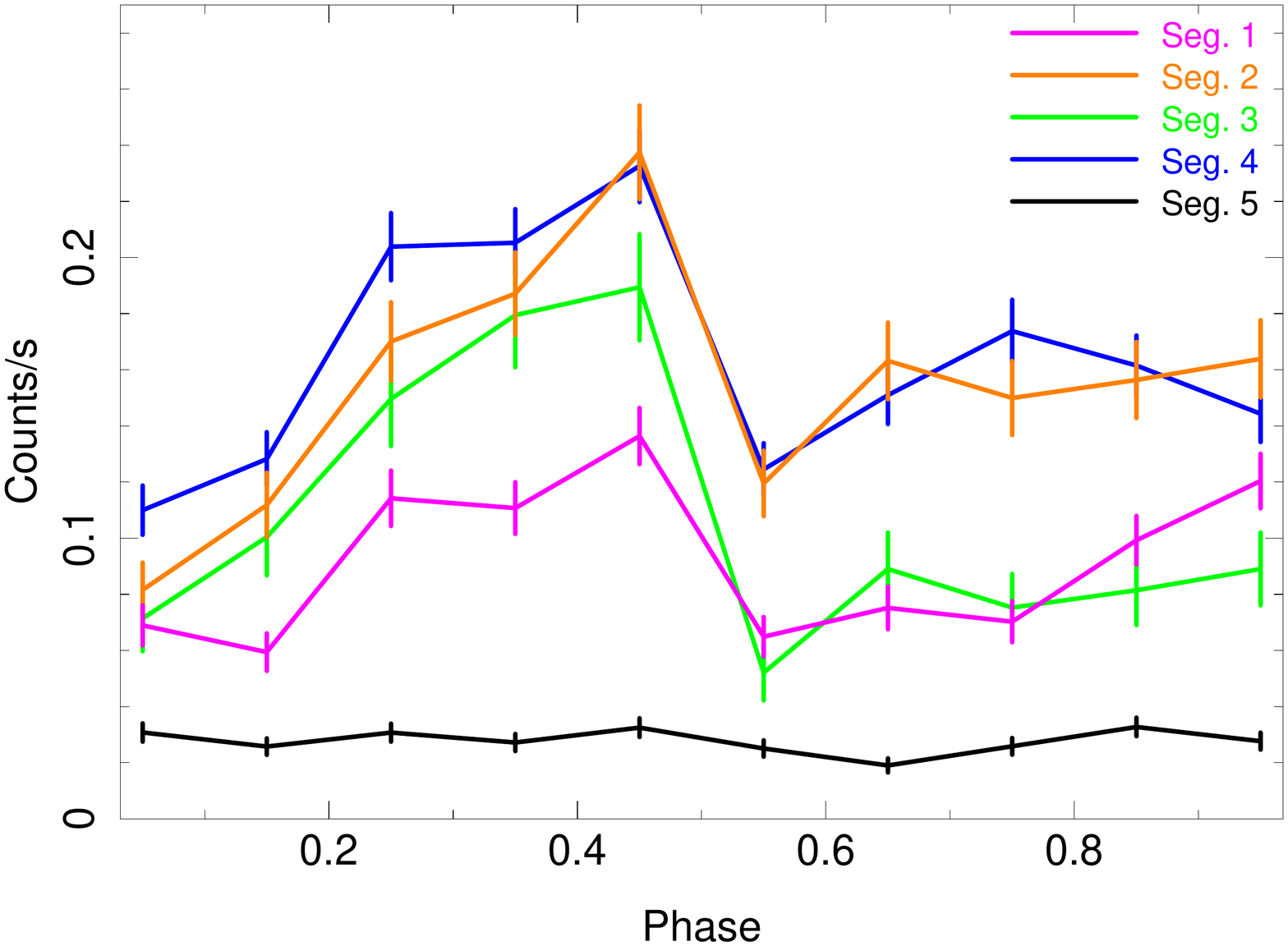}
    \caption{Pulse profiles of the continuum within 3--6\,keV (left panel) and 7.5--10\,keV (right panel) for Seg.\,1 to 5. For clarity, the profile of Seg.\,5 is multiplied by a factor of 12 in the left panel. Error bars indicate $1\sigma$ uncertainty. }
    \label{fig:pfold}
\end{figure*}

\label{sec:results}
The extracted light curves of the low-energy band (1--2.1\,keV), the Fe \Ka line (6.35--6.45\,keV), and the continua
within 3--6\,keV and 7.5--10\,keV are presented in Fig.~\ref{fig:open}.
A baseline continuum, estimated as 40\% of the count rate in the 6.0--6.2\,keV band, is subtracted from the Fe \Ka
fluxes. The coefficient of 40\% is estimated from the flux ratio between 6.35--6.45\,keV and 6.0--6.2\,keV from an
absorbed powerlaw model fitted to the spectrum of the observation.
All light curves are binned to 283.6\,s. The best pulse period estimated with the {\sc efsearch} program in FTOOLS
from the first 40 ks of the observation is 283.57\,s, consistent with the spin frequency measured with {\it Fermi}/GBM
(Fig.~\ref{fig:vx1_gbm}). As can be seen from Fig.~\ref{fig:open}, the continua within 3--6\,keV and 7.5--10\,keV 
show similar trends, while the trend of the low-energy band (1--2.1\,keV) is similar to that of the Fe \Ka line. 
The continua show two flares around 20\,ks and 35\,ks, and then they enter an extended low state from 42\,ks,
keeping quiescent until the end of the observation.
On the other hand, the Fe \Ka line reaches a high level during the flare around 35 ks, and then it drops a little and 
back to a high level later.
The flux ratios between the Fe \Ka and the continuum within 3--6\,keV are plotted in the bottom panel of Fig.~\ref{fig:open}.
The ratios with respect to the continuum within 7.5--10\,keV are similar and not presented.
The average Fe K$\alpha$/3--6\,keV ratios are 0.02 and 0.14, during the first 42 ks and thereafter, respectively.
Based on the flux levels of different bands, we divide the observation into five segments: 0-13.5 ks (1), 13.5-22.3 ks (2), 22.3-27.7 ks (3), 27.7-42.1 ks (4), and 42.1-72.2 ks (5). 

For each segment, we search for periodicity in the continua within 3--6\,keV and 7.5--10\,keV, the low-energy band 
(1--2.1\,keV), and the Fe \Ka line with the epoch folding method for light curves binned in 10\,s. The trial periods are between 100\,s and 400\,s, with a step of 0.1\,s and 20 phase bins. 
The significances of periodicity \citep[L-statistics;][]{Lea83,Lstat} of the trial periods are plotted 
in Fig.~\ref{fig:prd_search_5seg}. As can be seen, both the continua within 3--6\,keV and 7.5--10\,keV of 
Seg.\,1--4 show clear pulsations with a period $\sim283.6$\,s, while their periodicity is almost disappeared in Seg.\,5. On the other hand, as expected, the low-energy band and the Fe \Ka line show no apparent periodicity in all five segments.

To calculate the pulse profiles of the continuum, we fold the 3--6\,keV light curves of all five segments with the 
period of 283.57\,s with 20 phase bins. The resulting pulse profiles are plotted in the left panel of
Fig.~\ref{fig:pfold}, 
and for clarity, the pulse profile of Seg.\,5 is multiplied by a factor of 12. 
The profiles of Seg.\,1--4 show regular five-peaked behavior \citep{Rau90}, while the profile of 
Seg.\,5 is close to flat. The pulse profiles of the continuum within 7.5--10\,keV are plotted in 
the right panel of Fig.~\ref{fig:pfold}. They show a similar behavior as the continuum within 3--6\,keV, but with larger
uncertainties due to lower counts. For each segment, we calculate the pulsed fraction of the continuum, 
which is defined as $f=\frac{max(p)-min(p)}{max(p)+min(p)}$, where $p$ represents the value of pulse profile.
For Seg.\,1--5, the pulsed fractions of the continuum within 3--6\,keV are 0.42, 0.46, 0.53, 0.48, and 0.11, 
respectively, while the pulsed fractions within 7.5--10\,keV are 0.39, 0.49, 0.57, 0.36, and 0.26, respectively.

\section{Spectral results}
\begin{figure*}
    \includegraphics[angle=0,width=2\columnwidth]{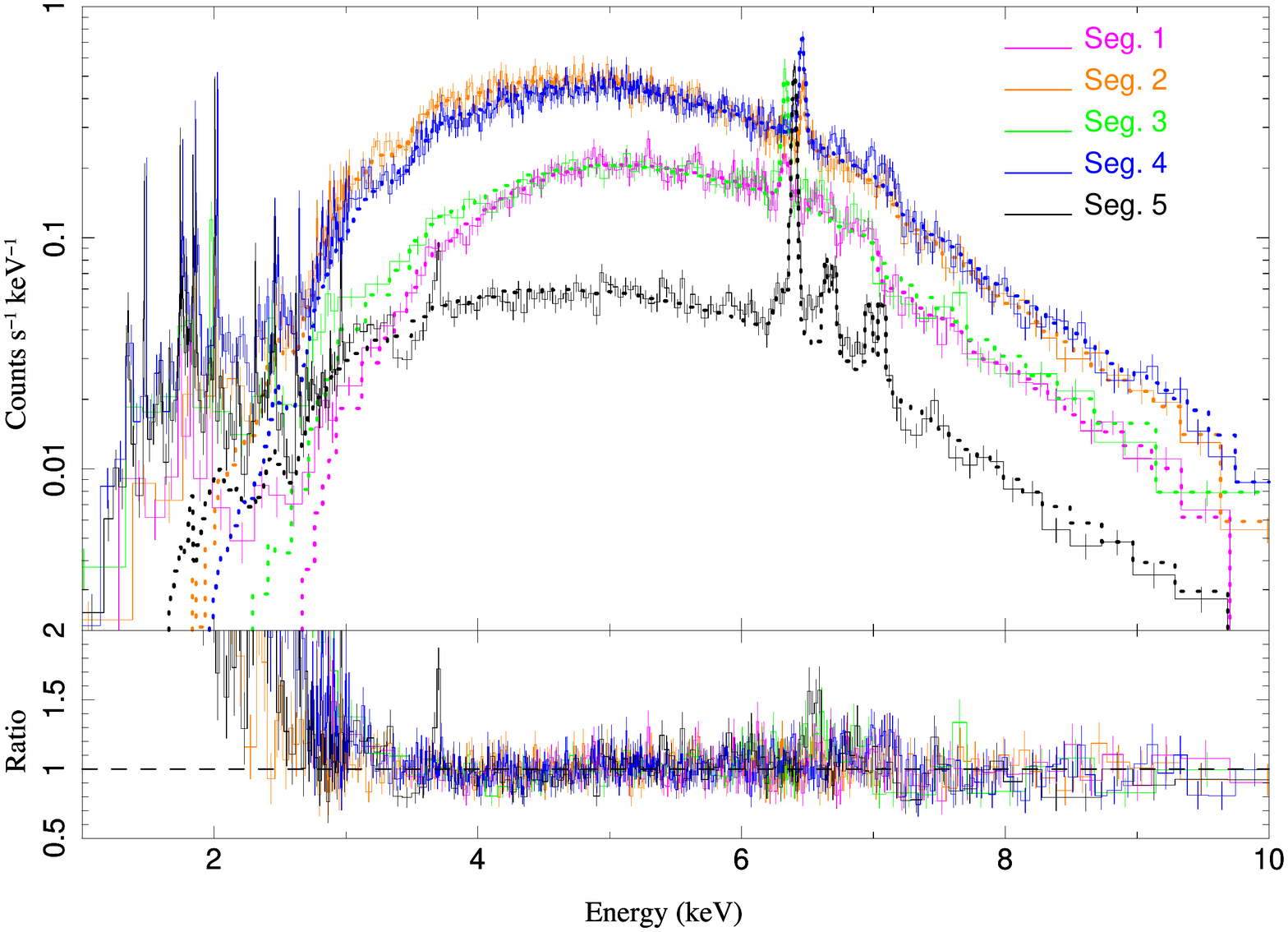}
    \caption{Extracted spectra of Vela X-1 for all five segments. For clarity, the spectra of Seg.\,1 and Seg.\,3 are redshifted by 0.01, and the spectra of Seg.\,2 and Seg.\,4 are blueshifted by 0.01. The histograms represent the observed data and the dotted lines represent the models fitted to the energy range of 3--10\,keV.}
    \label{fig:5seg_cont}
\end{figure*}
The extracted spectra of all five segments are plotted in Fig.~\ref{fig:5seg_cont}. 
They are binned with a piecewise minimum signal-to-ratio (S/N) of 5, 11, and 8 for energy ranges
of 1--3, 3--6, and 6--10\,keV, respectively. 
As can be seen, the continua of the two flares in Seg.\,2 and Seg.\,4 are clearly higher than other segments. The continuum drops significantly for Seg.\,5, while the continua of Seg.\,1 and 3 are in between. Meanwhile, the Fe \Ka
line is relatively low for the first two segments and starts to increase from Seg.\,3.
The drop of the continuum of Seg.\,5 makes its Fe \Ka line very prominent. There is a redward wing of the Fe \Ka line
of Seg.\,5, which is likely the Compton shoulder of the Fe \Ka line. The Fe \Kb line and highly-ionized \FeXXV and
\FeXXVI lines are also clearly seen in the spectrum of Seg.\,5. The spectrum of Seg.\,4 also shows similar highly-ionized Fe lines, but with a lower contrast due to a higher continuum level. Similar to the Fe \Ka line, the low-energy lines (1--2.1\,keV) of the low state of Seg.\,5 are as bright as those of Seg.\,4.

To quantify the spectral differences of different segments, we fit the spectra with a phenomenological model, 
consisting of an absorbed powerlaw and a Gaussian line (representing the Fe \Ka line). 
The centroid of Fe \Ka line has been fixed to 6.3995\,keV, which is the weighted value of neutral Fe K$\alpha_1$ and Fe K$\alpha_2$ \citep{KO1979}. For the spectrum of Seg.\,5, we add five additional Gaussian lines, representing Compton shoulder of Fe \Ka line, \FeXXV forbidden line, \FeXXV resonance line, \FeXXVI \Lya line, and Fe \Kb line respectively. 
As shown in previous studies \citep[e.g.][]{Schu2002,Wat2006,Grin2017}, the low-energy spectrum of Vela X-1 is composed of many photoionized lines and neutral fluorescence lines, indicating the coexistence of the cool and hot gas around the neutron star. 
We limit the fitting energy range to 3--10\,keV to avoid the complex low-energy lines.
The fitting results are listed in Table~\ref{tab:seg1-5} and \ref{tab:seg5_pl} and over-plotted in Fig.~\ref{fig:5seg_cont} as dotted lines. The absorption-corrected luminosity within 2--10\,keV and the Fe \Ka EW are also listed in Table~\ref{tab:seg1-5}. 
Note that the fitting is only aimed to quantify the observed spectral differences, not necessarily representing the intrinsic spectral properties. This is especially true for Seg.\,5, the emission of which should be dominated by scattered component.

From Table~\ref{tab:seg1-5}, we see that the continuum flux of the low state of Seg.\,5 dropped by a factor of
$10$ compared
to that of Seg.\,4. On the other hand, the Fe \Ka fluxes rise in Seg.\,3 and then remain at a
similar level. As a consequence, the Fe \Ka EW of Seg.\,5 is very high, $\sim0.6$\,keV, which is rarely seen 
during non-eclipse periods and is comparable to that of the eclipse period \citep{Wat2006}.

\begin{figure}
    \includegraphics[angle=0, width=\columnwidth]{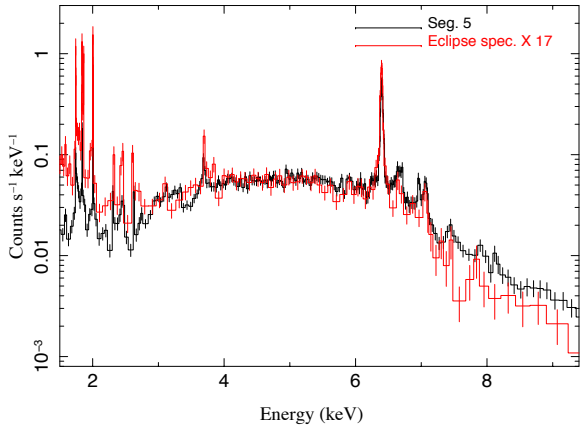}
    \caption{Comparison of the spectrum of the extended low state of Seg.\,5 (black) with
         the eclipse spectrum of Vela X-1 (red). The eclipse spectrum is multiplied by a factor of 17.
 Their overall shapes are quite similar.}
    \label{fig:seg5_ecli}
\end{figure}

To further illustrate the nature of the low state of Seg.\,5, in Fig.~\ref{fig:seg5_ecli}, we compare the eclipse
spectrum (extracted from ObsID 1926, observed by {\it Chandra}/HETG during the eclipse phase of Vela X-1) to
the spectrum of Seg.\,5. The eclipse spectrum is multiplied by a factor of 17.
It is interesting to see that their overall shapes are quite similar. Highly-ionized \FeXXV and \FeXXVI lines are not present in the eclipse spectrum. Below 2.5\,keV, the eclipse spectrum is relatively higher than the spectrum of Seg.\,5, indicating less low energy absorption of the eclipse spectrum.
Flux variations of Vela X-1 due to variable absorptions with a changing column density $\sim5\times10^{22}$
cm$^{-2}$ have been reported previously, and they could be caused by the unsteady accretion flow 
\citep[e.g.][]{Marti2014,Grin2017}. Apparently, the eclipse-like spectrum of Seg.\,5 is quite different from those 
variations caused by column density changes $\sim5\times10^{22}$ cm$^{-2}$.

\begin{table*}
    \renewcommand\arraystretch{1.5}
    \centering
	 \caption{Fitting results of an absorbed powerlaw plus a Gaussian line ($wabs\times powerlaw+zgauss$). }
    \begin{threeparttable}
    \begin{tabular}{ c c c c c c c c c c }
        \hline
        \multirow{2}{*}{\centering Seg.} & \multicolumn{3}{c}{Absorbed powerlaw} & \multicolumn{4}{c}{Fe \Ka line} & \multirow{2}{*}{\makecell{Luminosity\tnote{3}\\($10^{35}$ erg/s)}} & \multirow{2}{*}{$\chi^2_\nu$} \\
        \cmidrule{2-4}
        \cmidrule{5-8}
        ~ & $N_{\rm H} (10^{22}\,\rm cm^{-2}$) & $\Gamma$ & $N_1$\tnote{1} & $\sigma$ (eV) & $N_2$\tnote{2} & {\it z} ($10^{-4}$) & {\it EW} (eV) & ~ & ~ \\
        \hline
        1 & $27.5\pm1.3$ & $1.2\pm0.1$ & $0.22\pm0.06$ & $21.5\pm10.0$ & $5.7\pm1.7$ & $9.3\pm15.0$ & $37.0\pm11.0$ & $11.4\pm1.7$ & 1.01 \\
        2 & $14.3\pm0.9$ & $1.3\pm0.1$ & $0.40\pm0.08$ & $4.2\pm4.0$ & $9.2\pm2.2$ & $-6.0\pm6.8$ & $34.8\pm8.3$ & $16.5\pm2.3$ & 0.93 \\
        3 & $17.4\pm1.8$ & $0.5\pm0.2$ & $0.06\pm0.02$ & $17.4\pm5.6$ & $22.2\pm3.5$ & $3.2\pm6.6$ & $148.0\pm23.3$ & $7.6\pm0.8$ & 1.49 \\
        4 & $15.5\pm0.7$ & $1.0\pm0.1$ & $0.25\pm0.04$ & $12.6\pm3.3$ & $24.1\pm2.4$ & $5.4\pm3.5$ & $86.6\pm8.6$ & $15.8\pm1.0$ & 1.29\\
        5 & $5.7\pm1.2$ & $0.14\pm0.13$ & $0.005\pm0.001$ & $11.1\pm1.6$ & $22.7\pm1.2$ & $-1.1\pm1.6$ & $588.6\pm31.1$ & $1.59\pm0.1$ & 2.26\\
        \hline
    \end{tabular}
    \begin{tablenotes}
        \footnotesize
        \item[1] Normalization of the powerlaw, in units of $\rm photons\,keV^{-1}\,cm^{-2}\,s^{-1}$ at 1\,keV.
        \item[2] Normalization of the Gaussian line, in units of $10^{-4}\rm\,photons\,cm^{-2}\,s^{-1}$.
        \item[3] 2--10\,keV absorption-corrected luminosity.
    \end{tablenotes}
    \end{threeparttable}
    \label{tab:seg1-5}
\end{table*}

\begin{table*}
    \renewcommand\arraystretch{1.5}
    \centering
	\caption{Additional Gaussian lines ({\it zgauss} model, fixing redshift to 0) fitted to the spectrum of Seg.\,5.}
    \begin{threeparttable}
    \begin{tabular}{ccccccc}
    \hline
     & Fe \Ka Compton shoulder & \FeXXV F[z] & \FeXXV R[w] & \FeXXVI \Lya & Fe \Kb \\
    \hline
    {\it N}\tnote{1} & $2.2\pm0.7$ & $3.0\pm0.6$ & $2.9\pm0.6$ & $2.9\pm0.7$ & $3.5\pm0.8$ \\
    {\it E}\,(keV) & $6.29\pm0.02$ & $6.639\pm0.007$ & $6.701\pm0.007$ & $6.964\pm0.008$ & $7.064\pm0.007$ \\
    $\sigma$\,(eV)\tnote{2} & 40 & 10 & 10 & 10 & 10 \\
    \hline
    \end{tabular}
    \begin{tablenotes}
        \footnotesize
        \item[1] Normalization of the Gaussian line, in units of $10^{-4}\rm\,photons\,cm^{-2}\,s^{-1}$.
        \item[2] All the line widths in this table have been fixed during fitting.
    \end{tablenotes}
    \end{threeparttable}
    \label{tab:seg5_pl}
\end{table*}

\section{Discussion and conclusion}
\label{sec:con}
At the onset of an unusual spin-up period of Vela X-1, we have found an extended low state at orbital phases of 0.16--0.2, lasting for at least 30 ks, until the end of the observation. 
During this low state, the continuum pulsation is almost disappeared, and the Fe \Ka line has an eclipse-like EW of $0.6$\,keV. Highly-ionized \FeXXV and \FeXXVI lines are also very prominent in the low state. 
These highly-ionized lines were already present during the preceding flare period, but with a weaker contrast due to a stronger continuum level. 
The observed spectrum below 2\,keV is dominated by emission lines, which are also prominent in the low state, similar to the Fe \Ka line.

The phenomena of low states with pulsation cessation of Vela X-1 have been reported previously \citep[e.g.][]{Hay84,Ino84,Lap92,Krey1999}. 
The extended low state we found here is similar to that presented by \citet{Kret2000}. 
They reported a complete low state (lasting for $\sim15$\,ks), during which the pulsation ceased, and significant non-pulsed emission remained. 
Thanks to the wide energy range of RXTE (2--60\,keV), they found that the pulsation cessation of high energy photons (above 10\,keV) was less prominent than low energy photons, 
and the pulsation of high energy photons recovered earlier than low energy photons. 
They proposed a scenario that the pulsar was temporarily obscured by a thick clump ($N_{\rm H}\sim2\times10^{24}\rm cm^{-2}$), which was thought to be on a spatial scale of $10^{13}$\,cm (283\,lt-s) to destroy the coherence. 
We note that this scale is $\sim3$ times the binary separation of Vela X-1, much larger than the typical clump sizes (about the Sobolev length, $\sim0.01R_*$) found in recent wind simulations \citep[e.g.][]{Sund2018}. 

Some of the observed low states of Vela X-1 \citep{Krey2008} were explained as results of the propeller effect
\citep{IS75}. The unabsorbed continuum luminosity (2--10\,keV) of the low state discovered here is
$\sim1.6\rm\times10^{35}erg\,s^{-1}$, far beyond that required by the propeller effect for Vela X-1 \citep[$\sim 6
\times 10^{32} \rm \,erg\,s^{-1}$,][]{Krey2008}, indicating that it can not be due to the propeller effect. The similar Fe \Ka fluxes between the low state and the preceding flare period indicate that the intrinsic radiation illuminating the Fe K$\alpha$-emitting material during the low state should be similar to that of the preceding flare period.

The similarity of the spectral shapes of the low state and the eclipse phase (Fig.~\ref{fig:seg5_ecli}) points to a blocking scenario: 
the intrinsic pulsed X-ray emission is blocked by a thick gas structure, and the observed X-ray emission 
is mainly due to scattered/reflected emission. 
To further discuss the blocking scenario by a thick gas structure, it is helpful to know the distance of the Fe K$\alpha$-emitting gas ($R_{\rm Fe K\alpha}$) around the neutron star. 
Because the observed Fe \Ka fluxes during the eclipse period reduced significantly \citep[5--10\%,][]{Oha84,Wat2006}, the majority of the Fe \Ka fluxes should originate from a region smaller than the radius of the donor star ($70$\,lt-s). 
From the observed Fe \Ka fluxes within a few hours before the egress, \citet{Oha84} inferred a limit of 
$R_{\rm Fe K\alpha}<17$\,lt-s. 

A natural spatial scale of the Fe K$\alpha$-emitting region is the accretion radius, within which the wind material
can be accreted by the neutron star: $R_{\rm acc}=2GM/v_{\rm rel}^2$, where $v_{\rm rel}$ is the relative velocity of
the neutron star with respect to the wind material. 
The estimated masses of the neutron star of Vela X-1 range from 1.5 to 2.2${\rm M_\odot}$
\citep[e.g.][]{Quain2003, Rawls2011, Koenig2012, Fala2015}. We adopt a mass of $1.8{\rm M_\odot}$ with an uncertainty
about 20\%. Then one get $R_{\rm acc}\approx 10$\,lt-s for $v_{\rm rel}\sim400$ km\,s$^{-1}$ \citep[][]{GiGa2017}.
We looked for the time delay between the Fe \Ka line and the continuum (3--6\,keV) with the cross-correlation function 
for light curves in Seg.\,4 and found a weak peak around $5$\,lt-s, similar to the accretion radius 
calculated above. 

We note that as long as the thick gas structure is quasi-axisymmetric with respect to the rotational axis of the
neutron star, the illuminating radiation on the gas structure would be similar for any spin phases,
and the scattered emission will show no apparent pulsation. The gas structure does not need to be as large as the light travel distance of the spin period ($283$\,lt-s) to destroy the coherence.
Since the continuum within 7.5--10\,keV in the low state shows no apparent pulsation, the column density of the thick
gas should be larger than $10^{24}$\,cm$^{-2}$ to obscure 7.5--10\,keV photons significantly. The high Fe \Ka EW of
the low state also requires a column density of $N_{\rm H}\geq10^{24} \rm cm^{-2}$ \citep[for a thin spherical shell,
$EW \sim$ 0.3\,keV$\times$$N_{\rm H}/10^{24}$cm$^{-2}$,][]{Kal04}. 
Such a high column density is an order of magnitude greater than the average column density observed 
at orbital phases of 0.16--0.2 \citep[e.g.][]{Dor13}. There are also examples of column densities 
$\sim5\times10^{23}$ cm$^{-2}$ around these phases \citep[e.g.][]{Nag86,Marti2014}.
If the intrinsic X-ray emission is blocked by a thick gas clump on a spatial scale of 283 lt-s, it will also block the Fe \Ka line and the low-energy photons, which is not observed.
Therefore, it is most likely that a quasi-axisymmetric thick gas structure on a scale of the accretion radius blocks the intrinsic X-ray emission and produces the observed scattering-dominated X-ray emission in the low state.

As shown in Fig.~\ref{fig:vx1_gbm}, there is a sharp spin-down event preceding the spin-up event.
The spin-down rate fitted to the spin frequencies within MJD 57740--57774 is $(-1.5\pm0.8)\times10^{-13}\rm s^{-2}$, while the spin-up
rate fitted within MJD 57807--57831 is $(2.3\pm1.3)\times10^{-13}\rm s^{-2}$.
\citet{Shaku2012} proposed a model of subsonic settling accretion, which occurs when the plasma remains hot until it meets the magnetospheric boundary. This model works at X-ray luminosities below $4\times10^{36}{\rm erg\,s^{-1}}$ and predicts a positive correlation between the torque and the luminosity.
The average pulsed flux of the spin-down period (MJD 57740--57774) is $0.9\pm0.3\,\rm keV\,cm^{-2}\,s^{-1}$, similar to that of the spin-up period (MJD 57807--57831), $1.2\pm0.2\,\rm keV\,cm^{-2}\,s^{-1}$.
This is inconsistent with the prediction of the subsonic accretion model. Furthermore, the hot shell of the subsonic accretion model is optically-thin, with a Thompson optical depth $\sim0.03$, corresponding to a column density 
of $4.5\times10^{22} \rm cm^{-2}$. Such a column density is too low to explain the low state.

Next we discuss whether a standard thin disk model can explain the measured spin-up rate of the spin-up event. 
As shown in numerical simulations \citep{Mel19}, the thickness of formed disk depends on the cooling efficiency.
So one should keep
in mind that a thin disk model may not be applicable and the angular momentum transport might be more complex.
Following \citet{Bild1997}, we assume that a thin disk deposits its angular momentum at the magnetosphere and
all its angular momentum is transported to the pulsar. The spin-up torque 
$N\approx\dot{M}\sqrt{GM_{\rm X}r_{\rm m}}$,
where $M_{\rm X}$ is the pulsar mass and the magnetosphere radius $r_{\rm m}=\xi r_{\rm A}$, is a fraction of 
the Alfv{\'e}n radius. $\xi$ ranges from 0.5 to 1.
The accretion rate $\dot{M} =L_{\rm X} R_{\rm X}/G M_{\rm X}$, 
where $L_{\rm X}$ is the accretion luminosity and $R_{\rm X}$ is the pulsar radius. 
The Alfv{\'e}n radius is:
\begin{equation}
\label{eq:alfven}
    r_{\rm A} =\left(\frac{\mu^4}{2GM_{\rm X}\dot{M}^2}\right)^\frac{1}{7}
\end{equation}
where $\mu\simeq0.5B\,R_{\rm X}^3$ is the magnetic moment of a dipole-like magnetic field of the neutron star, with $B$ the surface field strength. 
The neutron star will spin-up at a rate: 
\begin{equation}
\label{eq:nudot}
\begin{split}
	\dot{\nu}=\frac{N}{2\pi I} & \simeq2.2\times10^{-13}{\rm s}^{-2}\left(\frac{L_{\rm X}}{10^{36} \rm erg \, s^{-1}}\right)^\frac{6}{7}\\ 
	& \times\left(\frac{B}{10^{12}\rm G}\right)^\frac{2}{7}\left(\frac{10\rm km}{R_{\rm X}}\right)^\frac{2}{7}\left(\frac{1.4 \rm M_\odot}{M_{\rm X}}\right)^\frac{10}{7}\xi^\frac{1}{2},
\end{split}
\end{equation}
where $I\simeq0.4M_{\rm X} R_{\rm X}^2$ is the pulsar's moment of inertia.
Because the spectrum of Seg.\,5 is dominated by scattered emission, one can not obtain the intrinsic
spectrum to infer the luminosity of Seg.\,5. So we estimate the luminosity based on the fitted spectrum of Seg.\,4.
We obtain a luminosity of $\sim7.6\times10^{36}$ \rm erg\,s$^{-1}$, adopting a cutoff energy of 30\,keV and a fold energy of 10\,keV \citep[][]{Krey1999}.
It leads to a spin-up rate of $\dot{\nu}\approx 9.6 \times 10^{-13} \rm s^{-2}$ for $B=2.6\times10^{12}\rm G$, $R_{\rm
X}=10 \rm km$, $M_{\rm X}=1.8 \rm M_\odot$, and $\xi=0.7$. 
The uncertainty of the estimation mainly comes from $L_X$, $M_X$, and $\xi$. 
Assuming all these factors have an uncertianty around 30\%, the estimated $\dot{\nu}$
should be accurate within a factor of 2.
Therefore, the observed spin-up rate seems to be on the same order of magnitude for a disk accretion scenario.
The preceding spin-down event could be due to accretion through a retrograde disk. 
Detailed simulations are needed to test whether it is possible to produce a transition from a retrograde disk 
to a prograde disk, with both disks lasting for tens of days.
We conclude that the observed continuum pulsation cessation and the high Fe \Ka EW of Seg.\,5 are consistent 
with the existence of an accretion disk, which blocks the intrinsic X-ray emission and leads to the following 
spin-up event.

The required column density of $N_{\rm H}\geq10^{24}{\rm cm}^{-2}$ and an emitting region of $5$\,lt-s imply an average density of $10^{13}{\rm cm}^{-3}$. 
This density is $\sim800$ times higher than the wind density at the location of the neutron star \citep{Sander2018}, implying that the wind material should be very condensed to form a disk. 
The neutral-like Fe K$\alpha$ line, with a centroid close to 6.4\,keV, implies relatively cool and low-ionized gas. These results indicate that the shocked material sould be cooled efficiently.

The existence of an accretion disk was also reported for other SgXBs, such as OAO 1657-415 \citep{Jenke2012, Taani2019}, GX 301-2 \citep{Nabi2019}, and Cen X-3 \citep{Tsu1996}. 
These results show that disk accretion is happening, although not always, in wind-fed SgXBs. 
A new mass transfer mode of wind-RLOF accretion was proposed in studies of symbiotic binaries by \citet{MP07,MP12}. \citet{Mellah2019} has proposed it as a possible explanation for ULXs hosting a neutron star. 
Our results show that disk accretion does occur in Vela X-1, but it is not a continuous process.
There are other physical factors that control the formation of disk. Further studies are needed to 
understand what is the dominant mechanism that leads to the formation and destruction of disk accretion.

\section*{Acknowledgements}
We thank our referee for constructive comments that improved the paper.
JL acknowledges the support by National Natural Science Foundation of China (NSFC, 11773035). 
LG acknowledges the supported by the National Program on Key Research and Development Project (2016YFA0400804), and by the NSFC (U1838114), and by the Strategic Priority Research Program of the Chinese Academy of Sciences (XDB23040100).
This research used data obtained from the \cha Data Archive.




\bibliographystyle{mnras}
\bibliography{vx0103}

\begin{thebibliography}{}
\makeatletter
\relax
\def\mn@urlcharsother{\let\do\@makeother \do\$\do\&\do\#\do\^\do\_\do\%\do\~}
\def\mn@doi{\begingroup\mn@urlcharsother \@ifnextchar [ {\mn@doi@}
  {\mn@doi@[]}}
\def\mn@doi@[#1]#2{\def\@tempa{#1}\ifx\@tempa\@empty \href
  {http://dx.doi.org/#2} {doi:#2}\else \href {http://dx.doi.org/#2} {#1}\fi
  \endgroup}
\def\mn@eprint#1#2{\mn@eprint@#1:#2::\@nil}
\def\mn@eprint@arXiv#1{\href {http://arxiv.org/abs/#1} {{\tt arXiv:#1}}}
\def\mn@eprint@dblp#1{\href {http://dblp.uni-trier.de/rec/bibtex/#1.xml}
  {dblp:#1}}
\def\mn@eprint@#1:#2:#3:#4\@nil{\def\@tempa {#1}\def\@tempb {#2}\def\@tempc
  {#3}\ifx \@tempc \@empty \let \@tempc \@tempb \let \@tempb \@tempa \fi \ifx
  \@tempb \@empty \def\@tempb {arXiv}\fi \@ifundefined
  {mn@eprint@\@tempb}{\@tempb:\@tempc}{\expandafter \expandafter \csname
  mn@eprint@\@tempb\endcsname \expandafter{\@tempc}}}

\bibitem[\protect\citeauthoryear{{Bildsten} et~al.,}{{Bildsten}
  et~al.}{1997}]{Bild1997}
{Bildsten} L.,  et~al., 1997, \mn@doi [\apjs] {10.1086/313060}, \href
  {https://ui.adsabs.harvard.edu/abs/1997ApJS..113..367B} {113, 367}

\bibitem[\protect\citeauthoryear{{Boynton}, {Deeter}, {Lamb}  \&
  {Zylstra}}{{Boynton} et~al.}{1986}]{Boy1986}
{Boynton} P.~E.,  {Deeter} J.~E.,  {Lamb} F.~K.,   {Zylstra} G.,  1986, \mn@doi
  [\apj] {10.1086/164443}, \href
  {https://ui.adsabs.harvard.edu/abs/1986ApJ...307..545B} {307, 545}

\bibitem[\protect\citeauthoryear{{Canizares} et~al.,}{{Canizares}
  et~al.}{2005}]{Cani05}
{Canizares} C.~R.,  et~al., 2005, \mn@doi [\pasp] {10.1086/432898}, \href
  {https://ui.adsabs.harvard.edu/abs/2005PASP..117.1144C} {117, 1144}

\bibitem[\protect\citeauthoryear{{Chodil}, {Mark}, {Rodrigues}, {Seward}  \&
  {Swift}}{{Chodil} et~al.}{1967}]{Chod1967}
{Chodil} G.,  {Mark} H.,  {Rodrigues} R.,  {Seward} F.~D.,   {Swift} C.~D.,
  1967, \mn@doi [\apj] {10.1086/149312}, \href
  {https://ui.adsabs.harvard.edu/abs/1967ApJ...150...57C} {150, 57}

\bibitem[\protect\citeauthoryear{{Davidson} \& {Ostriker}}{{Davidson} \&
  {Ostriker}}{1973}]{DO73}
{Davidson} K.,  {Ostriker} J.~P.,  1973, \mn@doi [\apj] {10.1086/151897}, \href
  {https://ui.adsabs.harvard.edu/abs/1973ApJ...179..585D} {179, 585}

\bibitem[\protect\citeauthoryear{{Davies}}{{Davies}}{1990}]{Lstat}
{Davies} S.~R.,  1990, \mnras, \href
  {https://ui.adsabs.harvard.edu/abs/1990MNRAS.244...93D} {244, 93}

\bibitem[\protect\citeauthoryear{{Deeter}, {Boynton}, {Lamb}  \&
  {Zylstra}}{{Deeter} et~al.}{1989}]{Dee1989}
{Deeter} J.~E.,  {Boynton} P.~E.,  {Lamb} F.~K.,   {Zylstra} G.,  1989, \mn@doi
  [\apj] {10.1086/167017}, \href
  {https://ui.adsabs.harvard.edu/abs/1989ApJ...336..376D} {336, 376}

\bibitem[\protect\citeauthoryear{{Doroshenko}, {Santangelo}, {Nakahira},
  {Mihara}, {Sugizaki}, {Matsuoka}, {Nakajima}  \& {Makishima}}{{Doroshenko}
  et~al.}{2013}]{Dor13}
{Doroshenko} V.,  {Santangelo} A.,  {Nakahira} S.,  {Mihara} T.,  {Sugizaki}
  M.,  {Matsuoka} M.,  {Nakajima} M.,   {Makishima} K.,  2013, \mn@doi [\aap]
  {10.1051/0004-6361/201321305}, \href
  {https://ui.adsabs.harvard.edu/abs/2013A&A...554A..37D} {554, A37}

\bibitem[\protect\citeauthoryear{{El Mellah}, {Sundqvist}  \& {Keppens}}{{El
  Mellah} et~al.}{2019a}]{Mellah2019}
{El Mellah} I.,  {Sundqvist} J.~O.,   {Keppens} R.,  2019a, \mn@doi [\aap]
  {10.1051/0004-6361/201834543}, \href
  {https://ui.adsabs.harvard.edu/abs/2019A&A...622L...3E} {622, L3}

\bibitem[\protect\citeauthoryear{{El Mellah}, {Sander}, {Sundqvist}  \&
  {Keppens}}{{El Mellah} et~al.}{2019b}]{Mel19}
{El Mellah} I.,  {Sander} A.~A.~C.,  {Sundqvist} J.~O.,   {Keppens} R.,  2019b,
  \mn@doi [\aap] {10.1051/0004-6361/201834498}, \href
  {https://ui.adsabs.harvard.edu/abs/2019A&A...622A.189E} {622, A189}

\bibitem[\protect\citeauthoryear{{Falanga}, {Bozzo}, {Lutovinov},
  {Bonnet-Bidaud}, {Fetisova}  \& {Puls}}{{Falanga} et~al.}{2015}]{Fala2015}
{Falanga} M.,  {Bozzo} E.,  {Lutovinov} A.,  {Bonnet-Bidaud} J.~M.,  {Fetisova}
  Y.,   {Puls} J.,  2015, \mn@doi [\aap] {10.1051/0004-6361/201425191}, \href
  {https://ui.adsabs.harvard.edu/abs/2015A&A...577A.130F} {577, A130}

\bibitem[\protect\citeauthoryear{{Gim{\'e}nez-Garc{\'\i}a}
  et~al.,}{{Gim{\'e}nez-Garc{\'\i}a} et~al.}{2016}]{GiGa2017}
{Gim{\'e}nez-Garc{\'\i}a} A.,  et~al., 2016, \mn@doi [\aap]
  {10.1051/0004-6361/201527551}, \href
  {https://ui.adsabs.harvard.edu/abs/2016A&A...591A..26G} {591, A26}

\bibitem[\protect\citeauthoryear{{Grinberg} et~al.,}{{Grinberg}
  et~al.}{2017}]{Grin2017}
{Grinberg} V.,  et~al., 2017, \mn@doi [\aap] {10.1051/0004-6361/201731843},
  \href {https://ui.adsabs.harvard.edu/abs/2017A&A...608A.143G} {608, A143}

\bibitem[\protect\citeauthoryear{{Hayakawa}}{{Hayakawa}}{1984}]{Hay84}
{Hayakawa} S.,  1984, \mn@doi [Advances in Space Research]
  {10.1016/0273-1177(84)90056-5}, \href
  {https://ui.adsabs.harvard.edu/abs/1984AdSpR...3...35H} {3, 35}

\bibitem[\protect\citeauthoryear{{Huenemoerder} et~al.,}{{Huenemoerder}
  et~al.}{2011}]{TGCat}
{Huenemoerder} D.~P.,  et~al., 2011, \mn@doi [\aj]
  {10.1088/0004-6256/141/4/129}, \href
  {https://ui.adsabs.harvard.edu/abs/2011AJ....141..129H} {141, 129}

\bibitem[\protect\citeauthoryear{{Illarionov} \& {Sunyaev}}{{Illarionov} \&
  {Sunyaev}}{1975}]{IS75}
{Illarionov} A.~F.,  {Sunyaev} R.~A.,  1975, \aap, \href
  {https://ui.adsabs.harvard.edu/abs/1975A&A....39..185I} {39, 185}

\bibitem[\protect\citeauthoryear{{Inoue}, {Ogawara}, {Ohashi}, {Waki},
  {Hayakawa}, {Kunieda}, {Nagase}  \& {Tsunemi}}{{Inoue} et~al.}{1984}]{Ino84}
{Inoue} H.,  {Ogawara} Y.,  {Ohashi} T.,  {Waki} I.,  {Hayakawa} S.,  {Kunieda}
  H.,  {Nagase} F.,   {Tsunemi} H.,  1984, \pasj, \href
  {https://ui.adsabs.harvard.edu/abs/1984PASJ...36..709I} {36, 709}

\bibitem[\protect\citeauthoryear{{Jenke} \& {Wilson-Hodge}}{{Jenke} \&
  {Wilson-Hodge}}{2017}]{Jenke2017}
{Jenke} P.~A.,  {Wilson-Hodge} C.~A.,  2017, in {Serino} M.,  {Shidatsu} M.,
  {Iwakiri} W.,   {Mihara} T.,  eds, 7 years of MAXI: monitoring X-ray
  Transients. p.~137

\bibitem[\protect\citeauthoryear{{Jenke}, {Finger}, {Wilson-Hodge}  \&
  {Camero-Arranz}}{{Jenke} et~al.}{2012}]{Jenke2012}
{Jenke} P.~A.,  {Finger} M.~H.,  {Wilson-Hodge} C.~A.,   {Camero-Arranz} A.,
  2012, \mn@doi [\apj] {10.1088/0004-637X/759/2/124}, \href
  {https://ui.adsabs.harvard.edu/abs/2012ApJ...759..124J} {759, 124}

\bibitem[\protect\citeauthoryear{{Joss} \& {Rappaport}}{{Joss} \&
  {Rappaport}}{1984}]{Joss1984}
{Joss} P.~C.,  {Rappaport} S.~A.,  1984, \mn@doi [\araa]
  {10.1146/annurev.aa.22.090184.002541}, \href
  {https://ui.adsabs.harvard.edu/abs/1984ARA&A..22..537J} {22, 537}

\bibitem[\protect\citeauthoryear{{Kallman}, {Palmeri}, {Bautista}, {Mendoza}
  \& {Krolik}}{{Kallman} et~al.}{2004}]{Kal04}
{Kallman} T.~R.,  {Palmeri} P.,  {Bautista} M.~A.,  {Mendoza} C.,   {Krolik}
  J.~H.,  2004, \mn@doi [\apjs] {10.1086/424039}, \href
  {https://ui.adsabs.harvard.edu/abs/2004ApJS..155..675K} {155, 675}

\bibitem[\protect\citeauthoryear{{Karino}, {Nakamura}  \& {Taani}}{{Karino}
  et~al.}{2019}]{Kar19}
{Karino} S.,  {Nakamura} K.,   {Taani} A.,  2019, \mn@doi [\pasj]
  {10.1093/pasj/psz034}, \href
  {https://ui.adsabs.harvard.edu/abs/2019PASJ...71...58K} {71, 58}

\bibitem[\protect\citeauthoryear{{Koenigsberger}, {Moreno}  \&
  {Harrington}}{{Koenigsberger} et~al.}{2012}]{Koenig2012}
{Koenigsberger} G.,  {Moreno} E.,   {Harrington} D.~M.,  2012, \mn@doi [\aap]
  {10.1051/0004-6361/201118397}, \href
  {https://ui.adsabs.harvard.edu/abs/2012A&A...539A..84K} {539, A84}

\bibitem[\protect\citeauthoryear{Krause \& Oliver}{Krause \&
  Oliver}{1979}]{KO1979}
Krause M.~O.,  Oliver J.,  1979, Journal of Physical and Chemical Reference
  Data, 8, 329

\bibitem[\protect\citeauthoryear{{Kretschmar} et~al.,}{{Kretschmar}
  et~al.}{1996}]{Kret1996}
{Kretschmar} P.,  et~al., 1996, \aaps, \href
  {https://ui.adsabs.harvard.edu/abs/1996A&AS..120C.175K} {120, 175}

\bibitem[\protect\citeauthoryear{{Kretschmar}, {Kreykenbohm}, {Wilms},
  {Staubert}, {Heindl}, {Gruber}  \& {Rothschild}}{{Kretschmar}
  et~al.}{2000}]{Kret2000}
{Kretschmar} P.,  {Kreykenbohm} I.,  {Wilms} J.,  {Staubert} R.,  {Heindl}
  W.~A.,  {Gruber} D.~E.,   {Rothschild} R.~E.,  2000, in {McConnell} M.~L.,
  {Ryan} J.~M.,  eds,  American Institute of Physics Conference Series Vol.
  510, American Institute of Physics Conference Series. pp 163--167 (\mn@eprint
  {arXiv} {astro-ph/9910539}), \mn@doi{10.1063/1.1303195}

\bibitem[\protect\citeauthoryear{{Kretschmar} et~al.,}{{Kretschmar}
  et~al.}{2019}]{Kret2019}
{Kretschmar} P.,  et~al., 2019, arXiv e-prints, \href
  {https://ui.adsabs.harvard.edu/abs/2019arXiv190508578K} {p. arXiv:1905.08578}

\bibitem[\protect\citeauthoryear{{Kreykenbohm}, {Kretschmar}, {Wilms},
  {Staubert}, {Kendziorra}, {Gruber}, {Heindl}  \& {Rothschild}}{{Kreykenbohm}
  et~al.}{1999}]{Krey1999}
{Kreykenbohm} I.,  {Kretschmar} P.,  {Wilms} J.,  {Staubert} R.,  {Kendziorra}
  E.,  {Gruber} D.~E.,  {Heindl} W.~A.,   {Rothschild} R.~E.,  1999, \aap,
  \href {https://ui.adsabs.harvard.edu/abs/1999A&A...341..141K} {341, 141}

\bibitem[\protect\citeauthoryear{{Kreykenbohm} et~al.,}{{Kreykenbohm}
  et~al.}{2008}]{Krey2008}
{Kreykenbohm} I.,  et~al., 2008, \mn@doi [\aap] {10.1051/0004-6361:200809956},
  \href {https://ui.adsabs.harvard.edu/abs/2008A&A...492..511K} {492, 511}

\bibitem[\protect\citeauthoryear{{Lapshov}, {Sunyaev}, {Chichkov}, {Dremin},
  {Brandt}  \& {Lund}}{{Lapshov} et~al.}{1992}]{Lap92}
{Lapshov} I.,  {Sunyaev} R.,  {Chichkov} M.,  {Dremin} V.,  {Brandt} S.,
  {Lund} N.,  1992, Soviet Astronomy Letters, \href
  {https://ui.adsabs.harvard.edu/abs/1992SvAL...18...16L} {18, 16}

\bibitem[\protect\citeauthoryear{{Leahy}, {Darbro}, {Elsner}, {Weisskopf},
  {Sutherland}, {Kahn}  \& {Grindlay}}{{Leahy} et~al.}{1983}]{Lea83}
{Leahy} D.~A.,  {Darbro} W.,  {Elsner} R.~F.,  {Weisskopf} M.~C.,  {Sutherland}
  P.~G.,  {Kahn} S.,   {Grindlay} J.~E.,  1983, \mn@doi [\apj]
  {10.1086/160766}, \href
  {https://ui.adsabs.harvard.edu/abs/1983ApJ...266..160L} {266, 160}

\bibitem[\protect\citeauthoryear{{Mart{\'\i}nez-N{\'u}{\~n}ez}
  et~al.,}{{Mart{\'\i}nez-N{\'u}{\~n}ez} et~al.}{2014}]{Marti2014}
{Mart{\'\i}nez-N{\'u}{\~n}ez} S.,  et~al., 2014, \mn@doi [\aap]
  {10.1051/0004-6361/201322404}, \href
  {https://ui.adsabs.harvard.edu/abs/2014A&A...563A..70M} {563, A70}

\bibitem[\protect\citeauthoryear{{Mart{\'\i}nez-N{\'u}{\~n}ez}
  et~al.,}{{Mart{\'\i}nez-N{\'u}{\~n}ez} et~al.}{2017}]{Mar17}
{Mart{\'\i}nez-N{\'u}{\~n}ez} S.,  et~al., 2017, \mn@doi [\ssr]
  {10.1007/s11214-017-0340-1}, \href
  {https://ui.adsabs.harvard.edu/abs/2017SSRv..212...59M} {212, 59}

\bibitem[\protect\citeauthoryear{{McClintock} et~al.,}{{McClintock}
  et~al.}{1976}]{McC1976}
{McClintock} J.~E.,  et~al., 1976, \mn@doi [\apjl] {10.1086/182142}, \href
  {https://ui.adsabs.harvard.edu/abs/1976ApJ...206L..99M} {206, L99}

\bibitem[\protect\citeauthoryear{{Mohamed} \& {Podsiadlowski}}{{Mohamed} \&
  {Podsiadlowski}}{2007}]{MP07}
{Mohamed} S.,  {Podsiadlowski} P.,  2007, in {Napiwotzki} R.,  {Burleigh}
  M.~R.,  eds,  Astronomical Society of the Pacific Conference Series Vol. 372,
  15th European Workshop on White Dwarfs. p.~397

\bibitem[\protect\citeauthoryear{{Mohamed} \& {Podsiadlowski}}{{Mohamed} \&
  {Podsiadlowski}}{2012}]{MP12}
{Mohamed} S.,  {Podsiadlowski} P.,  2012, \mn@doi [Baltic Astronomy]
  {10.1515/astro-2017-0362}, \href
  {https://ui.adsabs.harvard.edu/abs/2012BaltA..21...88M} {21, 88}

\bibitem[\protect\citeauthoryear{{Nabizadeh}, {M{\"o}nkk{\"o}nen}, {Tsygankov},
  {Doroshenko}, {Molkov}  \& {Poutanen}}{{Nabizadeh} et~al.}{2019}]{Nabi2019}
{Nabizadeh} A.,  {M{\"o}nkk{\"o}nen} J.,  {Tsygankov} S.~S.,  {Doroshenko} V.,
  {Molkov} S.~V.,   {Poutanen} J.,  2019, \mn@doi [\aap]
  {10.1051/0004-6361/201936045}, \href
  {https://ui.adsabs.harvard.edu/abs/2019A&A...629A.101N} {629, A101}

\bibitem[\protect\citeauthoryear{{Nagase}, {Hayakawa}, {Sato}, {Masai}  \&
  {Inoue}}{{Nagase} et~al.}{1986}]{Nag86}
{Nagase} F.,  {Hayakawa} S.,  {Sato} N.,  {Masai} K.,   {Inoue} H.,  1986,
  \pasj, \href {https://ui.adsabs.harvard.edu/abs/1986PASJ...38..547N} {38,
  547}

\bibitem[\protect\citeauthoryear{{Ohashi} et~al.,}{{Ohashi}
  et~al.}{1984}]{Oha84}
{Ohashi} T.,  et~al., 1984, \pasj, \href
  {https://ui.adsabs.harvard.edu/abs/1984PASJ...36..699O} {36, 699}

\bibitem[\protect\citeauthoryear{{Quaintrell}, {Norton}, {Ash}, {Roche},
  {Willems}, {Bedding}, {Baldry}  \& {Fender}}{{Quaintrell}
  et~al.}{2003}]{Quain2003}
{Quaintrell} H.,  {Norton} A.~J.,  {Ash} T.~D.~C.,  {Roche} P.,  {Willems} B.,
  {Bedding} T.~R.,  {Baldry} I.~K.,   {Fender} R.~P.,  2003, \mn@doi [\aap]
  {10.1051/0004-6361:20030120}, \href
  {https://ui.adsabs.harvard.edu/abs/2003A&A...401..313Q} {401, 313}

\bibitem[\protect\citeauthoryear{{Raubenheimer}}{{Raubenheimer}}{1990}]{Rau90}
{Raubenheimer} B.~C.,  1990, \aap, \href
  {https://ui.adsabs.harvard.edu/abs/1990A&A...234..172R} {234, 172}

\bibitem[\protect\citeauthoryear{{Rawls}, {Orosz}, {McClintock}, {Torres},
  {Bailyn}  \& {Buxton}}{{Rawls} et~al.}{2011}]{Rawls2011}
{Rawls} M.~L.,  {Orosz} J.~A.,  {McClintock} J.~E.,  {Torres} M. A.~P.,
  {Bailyn} C.~D.,   {Buxton} M.~M.,  2011, \mn@doi [\apj]
  {10.1088/0004-637X/730/1/25}, \href
  {https://ui.adsabs.harvard.edu/abs/2011ApJ...730...25R} {730, 25}

\bibitem[\protect\citeauthoryear{{Sadakane}, {Hirata}, {Jugaku}, {Kondo},
  {Matsuoka}, {Tanaka}  \& {Hammerschlag-Hensberge}}{{Sadakane}
  et~al.}{1985}]{Sada1985}
{Sadakane} K.,  {Hirata} R.,  {Jugaku} J.,  {Kondo} Y.,  {Matsuoka} M.,
  {Tanaka} Y.,   {Hammerschlag-Hensberge} G.,  1985, \mn@doi [\apj]
  {10.1086/162791}, \href
  {https://ui.adsabs.harvard.edu/abs/1985ApJ...288..284S} {288, 284}

\bibitem[\protect\citeauthoryear{{Sander}, {F{\"u}rst}, {Kretschmar},
  {Oskinova}, {Todt}, {Hainich}, {Shenar}  \& {Hamann}}{{Sander}
  et~al.}{2018}]{Sander2018}
{Sander} A.~A.~C.,  {F{\"u}rst} F.,  {Kretschmar} P.,  {Oskinova} L.~M.,
  {Todt} H.,  {Hainich} R.,  {Shenar} T.,   {Hamann} W.~R.,  2018, \mn@doi
  [\aap] {10.1051/0004-6361/201731575}, \href
  {https://ui.adsabs.harvard.edu/abs/2018A&A...610A..60S} {610, A60}

\bibitem[\protect\citeauthoryear{{Schulz}, {Canizares}, {Lee}  \&
  {Sako}}{{Schulz} et~al.}{2002}]{Schu2002}
{Schulz} N.~S.,  {Canizares} C.~R.,  {Lee} J.~C.,   {Sako} M.,  2002, \mn@doi
  [\apjl] {10.1086/338862}, \href
  {https://ui.adsabs.harvard.edu/abs/2002ApJ...564L..21S} {564, L21}

\bibitem[\protect\citeauthoryear{{Shakura}, {Postnov}, {Kochetkova}  \&
  {Hjalmarsdotter}}{{Shakura} et~al.}{2012}]{Shaku2012}
{Shakura} N.,  {Postnov} K.,  {Kochetkova} A.,   {Hjalmarsdotter} L.,  2012,
  \mn@doi [\mnras] {10.1111/j.1365-2966.2011.20026.x}, \href
  {https://ui.adsabs.harvard.edu/abs/2012MNRAS.420..216S} {420, 216}

\bibitem[\protect\citeauthoryear{{Shapiro} \& {Lightman}}{{Shapiro} \&
  {Lightman}}{1976}]{SL76}
{Shapiro} S.~L.,  {Lightman} A.~P.,  1976, \mn@doi [\apj] {10.1086/154203},
  \href {https://ui.adsabs.harvard.edu/abs/1976ApJ...204..555S} {204, 555}

\bibitem[\protect\citeauthoryear{{Sundqvist}, {Owocki}  \& {Puls}}{{Sundqvist}
  et~al.}{2018}]{Sund2018}
{Sundqvist} J.~O.,  {Owocki} S.~P.,   {Puls} J.,  2018, \mn@doi [\aap]
  {10.1051/0004-6361/201731718}, \href
  {https://ui.adsabs.harvard.edu/abs/2018A&A...611A..17S} {611, A17}

\bibitem[\protect\citeauthoryear{{Taani}, {Karino}, {Song}, {Al-Wardat},
  {Khasawneh}  \& {Mardini}}{{Taani} et~al.}{2019}]{Taani2019}
{Taani} A.,  {Karino} S.,  {Song} L.,  {Al-Wardat} M.,  {Khasawneh} A.,
  {Mardini} M.~K.,  2019, \mn@doi [Research in Astronomy and Astrophysics]
  {10.1088/1674-4527/19/1/12}, \href
  {https://ui.adsabs.harvard.edu/abs/2019RAA....19...12T} {19, 012}

\bibitem[\protect\citeauthoryear{{Tsunemi}, {Kitamoto}  \& {Tamura}}{{Tsunemi}
  et~al.}{1996}]{Tsu1996}
{Tsunemi} H.,  {Kitamoto} S.,   {Tamura} K.,  1996, \mn@doi [\apj]
  {10.1086/176652}, \href
  {https://ui.adsabs.harvard.edu/abs/1996ApJ...456..316T} {456, 316}

\bibitem[\protect\citeauthoryear{{Wang}}{{Wang}}{1981}]{Wang81}
{Wang} Y.~M.,  1981, \aap, \href
  {https://ui.adsabs.harvard.edu/abs/1981A&A...102...36W} {102, 36}

\bibitem[\protect\citeauthoryear{{Watanabe} et~al.,}{{Watanabe}
  et~al.}{2006}]{Wat2006}
{Watanabe} S.,  et~al., 2006, \mn@doi [\apj] {10.1086/507458}, \href
  {https://ui.adsabs.harvard.edu/abs/2006ApJ...651..421W} {651, 421}

\bibitem[\protect\citeauthoryear{{van Kerkwijk}, {van Paradijs}, {Zuiderwijk},
  {Hammerschlag-Hensberge}, {Kaper}  \& {Sterken}}{{van Kerkwijk}
  et~al.}{1995}]{vanKer1995}
{van Kerkwijk} M.~H.,  {van Paradijs} J.,  {Zuiderwijk} E.~J.,
  {Hammerschlag-Hensberge} G.,  {Kaper} L.,   {Sterken} C.,  1995, \aap, \href
  {https://ui.adsabs.harvard.edu/abs/1995A&A...303..483V} {303, 483}

\makeatother
\end{thebibliography}







\bsp	
\label{lastpage}
\end{document}